\documentclass[lettersize,journal]{IEEEtran}

\makeatletter
\newcommand{\Rmnum}[1]{\expandafter\@slowromancap\romannumeral #1@}
\makeatother
\usepackage{tabu}
\usepackage{verbatim}
\usepackage{graphicx}
\usepackage{bm}
\usepackage{multicol}
\usepackage{setspace}
\usepackage{subfig}
\usepackage{float}
\usepackage{amsmath}
\usepackage{epstopdf}
\usepackage{fancyhdr} % 添加页眉页脚
\usepackage{indentfirst}
\usepackage{enumerate}
\usepackage{lipsum}
\usepackage{amssymb}
\usepackage{stfloats}
\usepackage{bm}
\usepackage{threeparttable}
\usepackage{CJK}
\usepackage[justification=centering]{caption}
\usepackage{makecell}
\usepackage{caption}
\usepackage{cases}
\usepackage{algorithm}
\usepackage{algpseudocode}
\usepackage{amsmath}
\usepackage{graphics}
\usepackage{subfig}
\usepackage{color}
\usepackage{epsfig}
\usepackage{cite}

\def\BibTeX{{\rm B\kern-.05em{\sc i\kern-.025em b}\kern-.08em
		T\kern-.1667em\lower.7ex\hbox{E}\kern-.125emX}}

\begin{document}
	\title{Resonant Beam Communications: A New Design Paradigm and Challenges}
	\author{Yuanming Tian,~\IEEEmembership{Student Member, IEEE}, Dongxu Li,~\IEEEmembership{Student Member, IEEE}, Chuan Huang,~\IEEEmembership{Member, IEEE}, Qingwen Liu,~\IEEEmembership{Senior Member, IEEE}, and Shengli Zhou,~\IEEEmembership{Fellow, IEEE}

\thanks{
	
	Yuanming Tian and Dongxu Li are with the Future Network of Intelligence Institute and the School of Science and Engineering, The Chinese University of Hong Kong, Shenzhen 518172, China (e-mail: yuanmingtian@link.cuhk.edu.cn; dongxuli@link.cuhk.edu.cn).
	
	Chuan Huang is with the School of Science and Engineering and the Future Network of Intelligence Institute, The Chinese University of Hong Kong, Shenzhen 518172, China (e-mail: huangchuan@cuhk.edu.cn).
	
	Qingwen Liu is with the College of Electronics and
	Information Engineering, Tongji University, Shanghai 201804, China (e-mail:qliu@tongji.edu.cn).
	
	Shengli Zhou is with the Department of Electrical and Computer
	Engineering, University of Connecticut, Storrs, CT 06250, USA (e-mail:
	shengli.zhou@uconn.edu).}
}
\maketitle

	\begin{abstract}
	Resonant beam communications (RBCom), which adopt oscillating photons between two separate retroreflectors for information transmission, exhibit potential advantages over other types of wireless optical communications (WOC). However, echo interference generated by the modulated beam reflected from the receiver affects the transmission of the desired information. To tackle this challenge, a synchronization-based point-to-point RBCom system is proposed to eliminate the echo interference, and the design for the transmitter and receiver is discussed. Subsequently, the performance of the proposed RBCom is evaluated and compared with that of visible light communications (VLC) and free space optical communications (FOC). Finally, future research directions are outlined and several implementation challenges of RBCom systems are highlighted.
	
	\end{abstract}
	
    
	\section{Introduction}

    Wireless optical communications (WOC) adopt visible, infrared, or ultraviolet optical carriers for information transmissions in unguided propagation media\cite{Survey-WOC}. Conventional WOC technologies mainly consist of two types of scenarios: visible light communications (VLC) and free space optical communications (FOC). VLC shows several advantages, such as accurate indoor localization of mobile devices and enhanced communication confidentiality \cite{LimVLC}, and FOC can utilize almost unlimited available optical bandwidth and thus supports very high transmission rate \cite{WOC}. Moreover, WOC offer several advantages such as no spectrum licensing requirements and low interference\cite{Survey-WOC,LimVLC,WOC}. Therefore, WOC have become important supplements to conventional wireless communications\cite{LimVLC}.
    
	VLC utilizes light emitting diodes (LEDs) as sources \cite{LimVLC}, and modulates the information bits by changing the light intensity. LEDs with a large luminous angle enable the mobile multi-user communications, while its narrow modulation bandwidth and low radiation energy limit the transmission rate \cite{LimVLC}. Existing VLC systems are mainly adopted for the short range scenarios such as indoor communications, e.g., a high-speed VLC system with a Si-substrate yellow LED was demonstrated in \cite{AppVLC}, which achieved a peak transmission data rate 3.764 Gbps with a communication distance 1.2 m. 
	
   FOC utilizes laser as source \cite{WOC}, and modulates the information bits by changing the amplitude, frequency, and phase of the laser. Laser with high power and strong directivity enables FOC to achieve very high transmission rate and good confidentiality. However, establishing laser communication link requires an acquisition, tracking, and pointing (ATP) system \cite{ATP}, which limits the mobility and multi-user access capability of FOC. Therefore, FOC is mainly adopted for the medium and long range scenarios, such as unmanned aerial vehicle (UAV) and inter-satellite communications \cite{Survey-WOC}. The FOC system implemented in \cite{AppFSO} achieved a transmission data rate 1.8 Gbps and a communication distance up to 45000 km.

   Resonant beam communications are emerging and promising WOC technique, first proposed in \cite{Resonant}, and exhibit several advantages such as high signal-to-noise ratio (SNR), non-mechnical mobility and multiple access \cite{Resonant,Echo}. Moreover, RBCom are mainly adopted for the short and medium range scenarios, such as indoor and ground-to-air communication. In the RBCom system, both the transmitter and receiver adopt retroreflectors to construct a spatially distributed laser resonator, and the gain mediums enhance the power of the photons in the resonator to form resonant beam. An echo interference elimination scheme was proposed in \cite{Echo} based on frequency-band shifting and optical filtering. However, the resonant beam carrying information bits is filtered out by the pass-band optical filter and thus is underutilized, which significantly attenuates the beam energy and thus plays down the effective SNR of the received signal.
   
   \begin{comment}
   	To overcome these shortcomings, this article proposes a synchronization-based echo interference elimination scheme for RBCom, which addresses the link level echo interference and multiple access issues, and compares its performances with VLC and FOC. Based on the proposed scheme, various multiple access technologies are studied and their performances are analyzed.
   \end{comment}

   To overcome these shortcomings, this article proposes a synchronization-based echo interference elimination scheme, which addresses the echo interference issues, discusses the design for the transmitter and receiver, and compares the proposed RBCom with the conventional VLC and FOC. Furthermore, future research directions are outlined and several challenges of implementing the RBCom systems are highlighted.
     
	\section{A New Design Paradigm}
	\label{Structure}

	This section commences the introduction of the resonant beam creation. Subsequently, two echo interference elimination schemes, utilizing distinct information-bearing methods, are proposed for the transmitter design. The receiver design is then formulated through the derivation of two corresponding demodulation methods. This proposed paradigm makes full utilization of the beam energy within the cavity, ensuring the effective transmission of the desired information while eliminating the echo interference.

	\subsection{Resonant Beam Creation}

\begin{figure}[htb]
	\centering
	\includegraphics[scale=0.45]{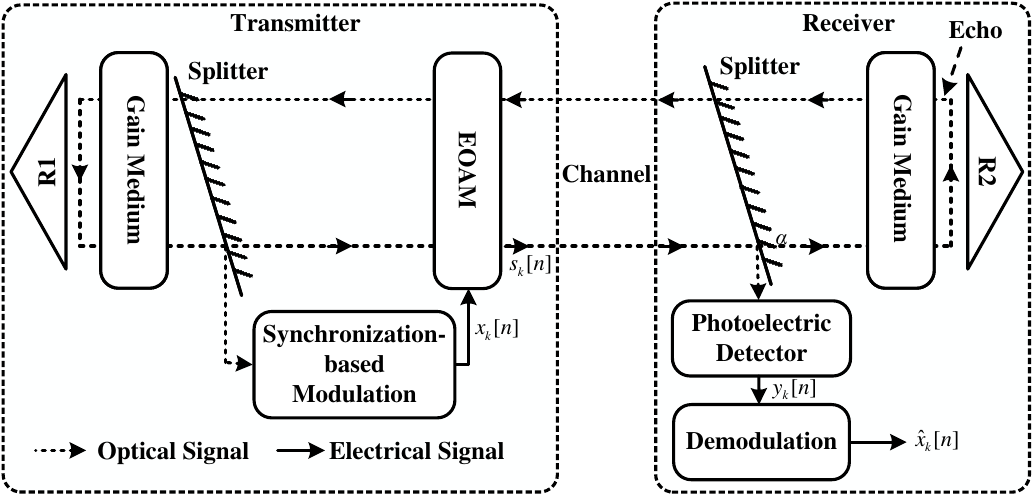}
	\caption{Proposed framework of point-to-point RBCom system.}
	\label{P2P structure}
    \end{figure}

	As depicted in Fig. \ref{P2P structure}, we propose a point-to-point RBCom system based on the distributed laser cavity first introduced by G. J. Linford in \cite{LongLaser}. Utilizing the property that retroreflectors reflect the beam back along the incident direction, two retroreflectors (R1 and R2) are placed at both the transmitter and the receiver, respectively, to construct an oscillating optical path composed of free space photons connecting the two ends. The gain medium has a finite gain bandwidth\footnote{Gain bandwidth refers to the full width at half maximum (FWHM) of the gain medium\cite{Gainmeidum}.} and a gain curve that approximately obeys a Lorentzian profile\footnote{Gain provided by the gain mediums takes its maximum value at the center frequency of the gain medium, and gradually decreases as the frequency moves away from the center frequency.}\cite{Gainmeidum}. In the cavity, gain mediums are equipped in front of the retrorelfectors, and they absorb the pump energy and amplify the passing beam. 
  
	During each reflection round\footnote{One reflection round means that the beam travels one cycle between the two separate retroreflectors, the duration of which can be calculated as the ratio of twice the distance between the two retrorelfectors to the speed of light.}, the beam moves cyclically between the transmitter and the receiver. When the gain provided by the gain mediums is greater than the link loss caused by the beam propagating on the optical path, the intensity of the beam increases. However, the intensity increment of one reflection round gradually decreases with the increase of the beam  intensity\cite{Gainmeidum}, until it reaches zero, i.e., the gain provided by the gain mediums is equal to the link loss. Then, the resonant beam reaches a steady state, where the beam intensity becomes constant over time \cite{PartOne}.

    \subsection{Transmitter Design}
   \label{Transmitter}

    At the transmitter, the resonant beam is first amplified by the gain medium, and then divided into two sub-beams by the splitter: one beam is sent to the synchronization-based modulation to eliminate the echo interference, which will be introduced later; the other one is sent to the electro-optical amplitude modulator (EOAM) to carry information symbols. The EOAM modulates the beam passing from the transmitter to the receiver by changing the intensity of the incident beam according to the voltage of the input electrical signal, and attenuates the intensity of the incident beam passing from the receiver to the transmitter\cite{EOAM}.    
     
    \begin{comment}
    The attenuated beam is absorbed by the EOAM, and the degree of attenuation is determined by the wavelength of beam and the type of EOAM, generally around 10$\%$.
    In this approach, the Electric-Opto-Amplitude Modulator (EOAM) modulates the beam traveling from the transmitter to the receiver by adjusting the intensity of the incident beam in response to the voltage of the input electrical signal. Conversely, the intensity of the incident beam passing from the receiver to the transmitter is attenuated by the EOAM \cite{EOAM}. 
    \end{comment}

    The RBCom system adopts a frame-based transmission mechanism, transmitting one frame of symbols in each reflection round. The number of symbols in each frame is determined by the modulation bandwidth of the EOAM and the distance between the transmitter and the receiver. Each frame consists of two parts: first part is the synchronization sequence (SS), which serves to realize symbol-level synchronization across different frames at the transmitter and the frame synchronization at the receiver; second part is the information sequence (IS), which is utilized for information transmission.
        
    We propose two innovative schemes to eliminate the echo interference, i.e., the direct and adaptive modulation schemes, each of which employs a distinct information-bearing method. 
   
    \subsubsection{Direct Modulation Scheme} 
\begin{figure}[htb]
    	\centering
    	\includegraphics[scale=0.42]{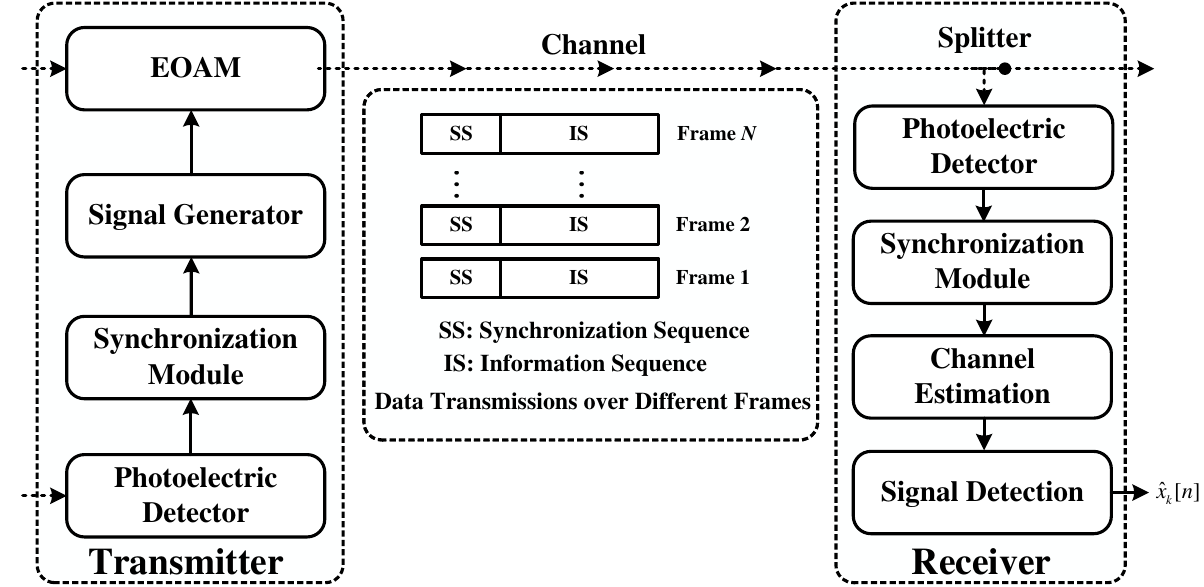}
    	\caption{Direct modulation echo interference elimination at the transmitter and receiver.}
    	\label{EIE}
    \end{figure}

In this scheme, the EOAM directly modulates the target signal onto the beam after the synchronization is completed, without consideration of the beam intensity. As depicted in Fig. \ref{EIE}, the information-bearing method consists of three steps. First, the photoelectric detector samples the beam, converts it into an electrical signal based on the beam intensity, and then inputs the signal to the synchronization module. Subsequently, the synchronization module performs real-time detection of the signal, and sends a notification signal to the signal generator upon detection of the SS. Finally, upon receipt of the notification signal, the signal generator generates an electrical signal carrying information symbols and feeds it to the EOAM. Here, we denote $x_k[n]$ as the $n$-th information symbol in frame $k$, and thus the $n$-th modulated information symbol in frame $k$ is given as 
    \begin{equation}
    	s_k[n]=\begin{cases}
    		\sqrt{P_t} x_k[n], & k=1, \\
    		\sqrt{h(s_{k-1}[n])}x_k[n], & k=2,3,\cdots, 
    	\end{cases} n=1,\cdots,N,
    \end{equation}
    where $P_t$ is the transmission power, and $h(s_{k-1}[n])$ is the channel coefficient as a function of the modulated information symbol $s_{k-1}[n]$ in frame $k-1$, and $N$ is the frame length.
    
    This scheme realizes the symbol-level multiplicative superposition across different frames, which results in mutual influence across frames and the transition probabilities of transmitted symbols vary with reflection rounds, making the RBCom channel an infinite-state Markov channel \cite{PartOne}. The channel coefficient is modeled as a link gain function with respect to the beam intensity, which consists of the gain provided by the gain mediums, the link loss, and the intensity attenuation caused by the optical devices. Specifically, the link gain function is given as 
    \begin{equation}
    	h(s_k[n])=\alpha\delta^2G_TG_Rs_{k}^2[n],
    \end{equation}
    where $\alpha$ is the transmission coefficient of the splitter, $\delta$ is the link loss between the transmitter and receiver, and $G_T$ and $G_R$ are the gain provided by the gain mediums at the transmitter and receiver, respectively.

    \subsubsection{Adaptive Modulation Scheme} In this scheme, the gain provided by the gain mediums is dynamically varied in response to the intensity of the modulated beam reflected from the receiver. To achieve this, the pump source is controlled in real-time based on the intensity of the beam in the transmitter. This approach effectively compensates the intensity of the beam, and ensures that the beam intensity at the input of the EOAM remain the same as that of the steady-state resonant beam, which achieves the elimination of the undesired echo interference. The EOAM then modulates the information symbols onto the resonant beam, realizing a finite-order intensity modulation\footnote{Finite-order intensity modulation is a technique that manipulates the intensity of the beam to transmit information, and has a finite number of modulation states\cite{Survey-WOC}.}. Here, the $n$-th modulated information symbol in frame $k$ is given as 
    \begin{equation}
    s_k[n] =\sqrt{P_t}x_k[n],k=1,2,\cdots.
    \end{equation}
    As a result, the RBCom channel is modeled as an additive white Gaussian noise (AWGN) channel, without mutual influence across frames. 

    \subsection{Receiver Design}
    \label{Receiver}

    The modulated beam propagates in the cavity and arrives at the receiver. The splitter divides the incident beam into two sub-beams: one beam is sent to the photoelectric detector, which converts the beam into an electrical signal and inputs it to the demodulation for detection of the information symbols based on the specific demodulation method; the other one is amplified by the gain medium, and then reflected back to the transmitter by R2 to maintain the resonance.
    
   In view of the different information-bearing methods employed by the two echo interference elimination schemes, we propose two demodulation methods to effectuate the detection of information symbols at the receiver.
    
    \subsubsection{Direct Modulation Scheme} In this scheme, the $n$-th received symbol in frame $k$ is given as 
    \begin{equation}
    \begin{split}
      y_k[n] &=\sqrt{(1-\alpha)\delta}s_k[n] + z_k[n]\\
             &=\begin{cases}
             	\sqrt{(1-\alpha)\delta P_t}x_k[n]+z_k[n],& k=1,\\
             	\sqrt{(1-\alpha)\delta h(s_{k-1}[n])}x_k[n]+z_k[n],& k=2,\cdots, 
             \end{cases}
    \end{split}
    \end{equation}
    where $z_k[n]$'s are independent and identically distributed (i.i.d.) additive AWGN with mean zero and variance $\sigma^2$.
    
    As depicted in Fig. \ref{EIE}, the demodulation method consists of three steps: First, the synchronization module identifies the SS by detecting the input signal in real-time, so as to realize the frame synchronization at the receiver. Subsequently, the SS are employed for channel estimation to derive an estimated link gain function $\widehat{h}(\cdot)$. Finally, the demodulation obtains the modulated information symbols of the previous reflection round by taking the ratio of the received symbols in the previous reflection round to the square root of the product of the link loss and the reflection coefficient of the splitter, i.e.,
     \begin{equation}
    	\widehat{s}_{k-1}[n] = \frac{y_k[n]}{\sqrt{(1-\alpha)\delta}},
    \end{equation}
    the estimated link gain function $\widehat{h}(\cdot)$ is then utilized to calculate channel coefficients that corresponds to the modulated information symbols, i.e, $\widehat{h}(\widehat{s}_{k-1}[n])$, and thus the desired information symbols of the current reflection round is derived by taking the ratio of the received symbols in the current reflection round to the square root of the product of the link loss, the reflection coefficient, and the corresponding channel coefficients of the previous reflection round. Hence, the $n$-th demodulated information symbol in frame $k$ is given as
    \begin{equation}
    	\widehat{x}_k[n]=\frac{y_k[n]}{\sqrt{(1-\alpha)\delta h(s_{k-1}[n])}}.
    \end{equation}
    This demodulation method ensures that the information symbols of each frame are detected without being affected by the echo interference.
   
    \subsubsection{Adaptive Modulation Scheme} In this scheme, the $n$-the received symbol in frame $k$ is given as
    \begin{equation}
    	y_k[n]=\sqrt{(1-\alpha)\delta}s_k[n] +z_k[n].
   \end{equation}
   This scheme adopts the same method as the direct modulation scheme to realize the frame synchronization at the receiver. As analyzed for the transmitter design, the information-bearing method is equivalent to a finite-order intensity modulation, which is a commonly used modulation method in FOC\cite{Survey-WOC}. Hence, the receiver directly obtains the detected symbols through energy detection\cite{Survey-WOC}.

    \subsection{Discussion}

   Two proposed echo interference elimination schemes have different advantages and disadvantages. Specifically, the direct modulation scheme implements the synchronization module and the signal generator through a field programmable gate array (FPGA), which results in an economical overall design and diminishes the challenges associated with its implementation. Nevertheless, the specific demodulation method requires channel estimation, which increases the detection complexity. In contrast, the adaptive modulation scheme adopts energy detection with lower detection complexity at the receiver, but poses significant challenges to the implementation at the transmitter  due to the stringent requirements on the response time of the pump source and the gain medium\cite{Gainmeidum}. Moreover, other optical devices such as retroreflectors, gain medium and EOAM, are highly integrated and easy to manufacture, so that the overall system design has moderate cost and complexity.

   \section{Comparisons}

   This section discusses and compares the performance of the RBCom, VLC, and FOC under the point-to-point communication scenario, taking into consideration key performance metrics such as transmission rate and two-way communication ability. Table \ref{Comparison one} compares the proposed RBCom with the conventional VLC and FOC.

\begin{table} [!htbp]
	\small
	\centering
	\caption{Performance comparisons of three optical communication technologies.}  
	\begin{spacing}{1.1}
		\begin{tabu}{p{3.5cm}<{\centering}|p{1.2cm}<{\centering}|p{1.2cm}<{\centering}|p{1.2cm}<{\centering}}
			\tabucline[1.5pt]{-}
			\textbf{Metric} & \textbf{RBCom}  & \textbf{VLC} & \textbf{FOC} \\
			\tabucline[1.5pt]{-}
			\hline  
			Transmission Rate & Average & Low & High \\
			\hline 
			Mobility & Average & Good & Bad \\
			\hline    
			Two-way Communication & Good & Bad & Average \\
			\hline
			Safety & Good & Good & Average\\ 
			\hline
			Confidentiality & Good & Average & Good  \\
			\hline 
			\tabucline[1.5pt]{-}
		\end{tabu}  
	\end{spacing}  
	\label{Comparison one}
\end{table}    

  \subsection{{Transmission}  Rate}
  Here, we consider a typical scenario to compare the transmission rate of the proposed RBCom with the direct modulation echo interference elimination scheme, VLC, and FOC through simulation. The setups for the scenario are given as follows: set the wavelength and the diffraction angle of the resonant beam and laser as $1064$ nm and 1.2$\times$10$^{-3}$ rad, respectively, the radius, length, and the saturation intensity of the gain medium as 3 mm, 1 cm, and 1.2$\times$10$^7$ W/m$^2$, respectively, the transmission coefficient of the splitter as 0.9, the half-intensity radiation angle and the irradiation angle of LED as 60$^{\circ}$ and 45$^{\circ}$, respectively, the semi-angle of the receiver's field-of-view as 90$^{\circ}$, the physical area of the photo-diode as 4 cm$^2$, the refractive index as 1.5, the gain of optical filter as 1, the optical to electric conversion efficiency as 53\%, the modulation bandwidth of the EOAM, LED, and laser as 100 MHz, the transmission power as 1 W, and the power spectral density (PSD) and bandwidth of noise as $-$170 dBm/Hz and 100 MHz, respectively.

  \begin{figure}[htb]
  	\centering
  	\subfloat[VLC vs. RBCom]{\includegraphics[scale=0.4]{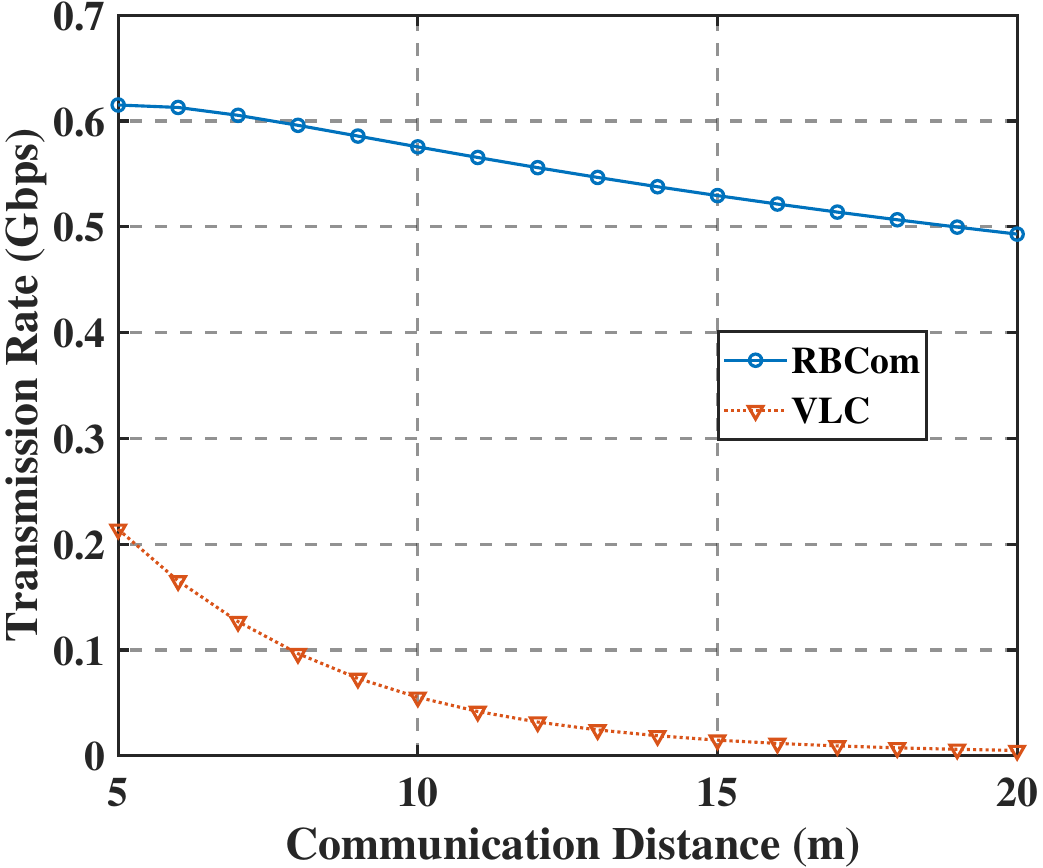}}\\
  	\subfloat[RBCom vs. FOC]{\includegraphics[scale=0.4]{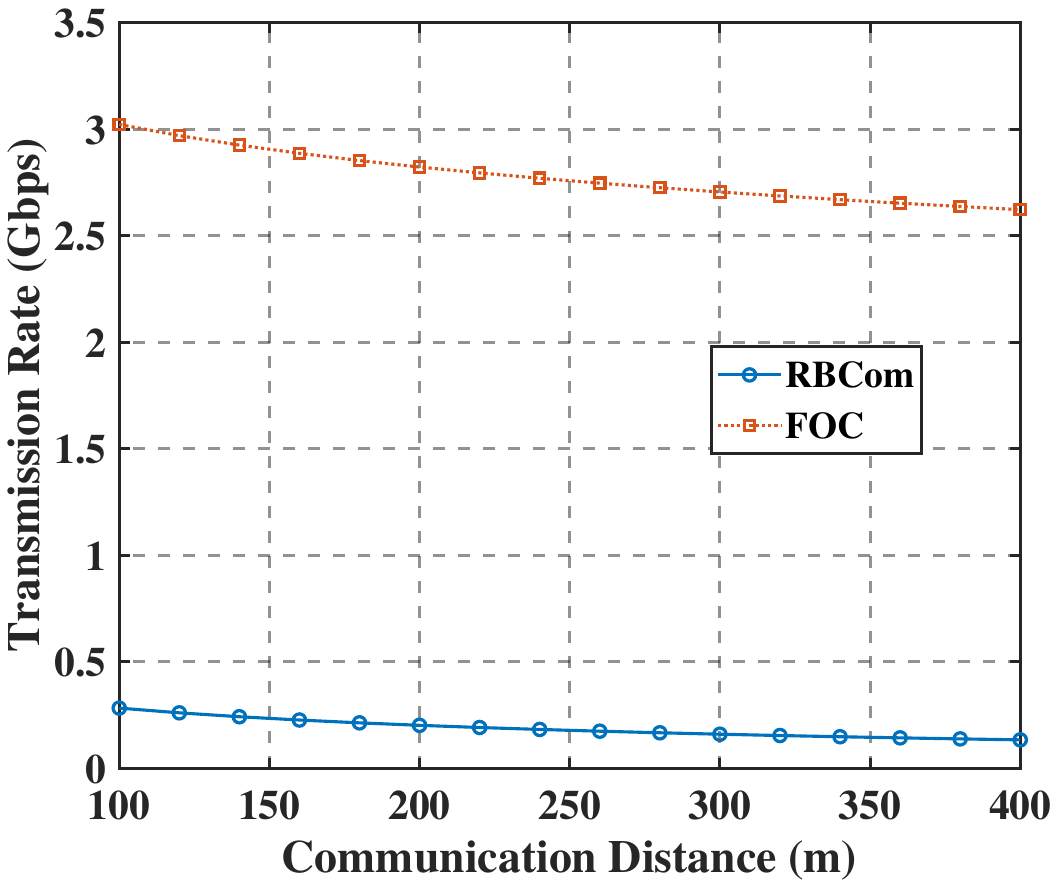}}
  	\caption{Comparisons of transmission rates of different WOC technologies.}
  	\label{Comparison_data}      
  \end{figure}

  Fig. \ref{Comparison_data} plots the transmission rates as functions of the communication distance between the transmitter and the receiver for the RBCom \cite{PartOne}, VLC \cite{VLCdatarate}, and FOC\cite{Survey-WOC}, respectively. From Fig. \ref{Comparison_data}(a), we observe that the transmission rate of RBCom is higher than that of VLC, since the energy of the resonant beam is more concentrated than that of LEDs, and the link loss of VLC significantly increases with the distance\cite{LimVLC}. From Fig. \ref{Comparison_data}(b), we observe that the transmission rate of the RBCom is lower than that of FOC,  disparity is attributed to the following three underlying reasons. Firstly, the splitter at the receiver divides the incident beam into two sub-beams. The beam utilized for information detection is allocated with only 10\% of the total power under the considered scenario, whereas the residual power is employed to maintain the resonance. Secondly, the Markov characteristics of the RBCom channel result in correlations between frames, which further reduces the transmission rate \cite{PartOne}. Finally, the peak power of the received signal is proportional to the square of the modulation depth \cite{PartOne}, which is limited by the gain capability of the gain mediums and thus much lower than that of FOC.

  \begin{comment}

  since the splitter at the receiver divides the incident beam into two sub-beams, and the power of the beam used to detect the information bits accounts for only 10\% of the total power. Moreover, the peak received signal power is limited by the amplitude modulation depth \cite{PartOne}, and the Markov characteristics of the RBCom channel lead to correlation between frames, further reducing the transmission rate \cite{PartOne}.
     \end{comment}

\begin{comment}
   \begin{figure}[htb]
   	\centering 
   	\subfigure[]{ 
   		%% label for first subfigure
   		\includegraphics[width=3.0in,height=2.7in]{figure//fig_light_RBC.pdf}
   	    %\label{detection}
   	}
   	\hspace{0.08in}
   	\subfigure[]{ 
   		%% label for second subfigure 
   		\includegraphics[width=3.0in,height=2.7in]{figure//fig_laser_RBC.pdf}
   	} 
   	\caption{Comparisons of transmission rate of different WOC technologies under the same transmission power. (a) VLC v.s. RBCom; (b) RBCom v.s. FOC.} 
   	\label{Comparison_da}
   \end{figure}
\end{comment}

   \subsection{Mobility}
   
   Based on the self-alignment property of the retroreflectors, an oscillating path is established between the transmitter and receiver when they relatively move, allowing them to maintain the connection at a low speed and with a direction angle\footnote{Direction angle refers to the angle between the moving direction of the receiver and the initial cavity.} of approximately 90$^{\circ}$ \cite{RBCcapacity}. Under the mobile scenario, the RBCom channel is affected by the propagation loss and the Doppler shift\cite{RBCcapacity}, which cause the communication link to break after a finite number of reflection rounds. To numerically analyze the mobility performance of the proposed RBCom, we adopt a typical scenario, where the initial distance between the transmitter and the receiver is set at 200 m, the transmission power is set at 1 W, and other parameters are the same as those described in Section III-A.
     
   \begin{figure}[htbp]
   	\centering
   	\includegraphics[scale=0.4]{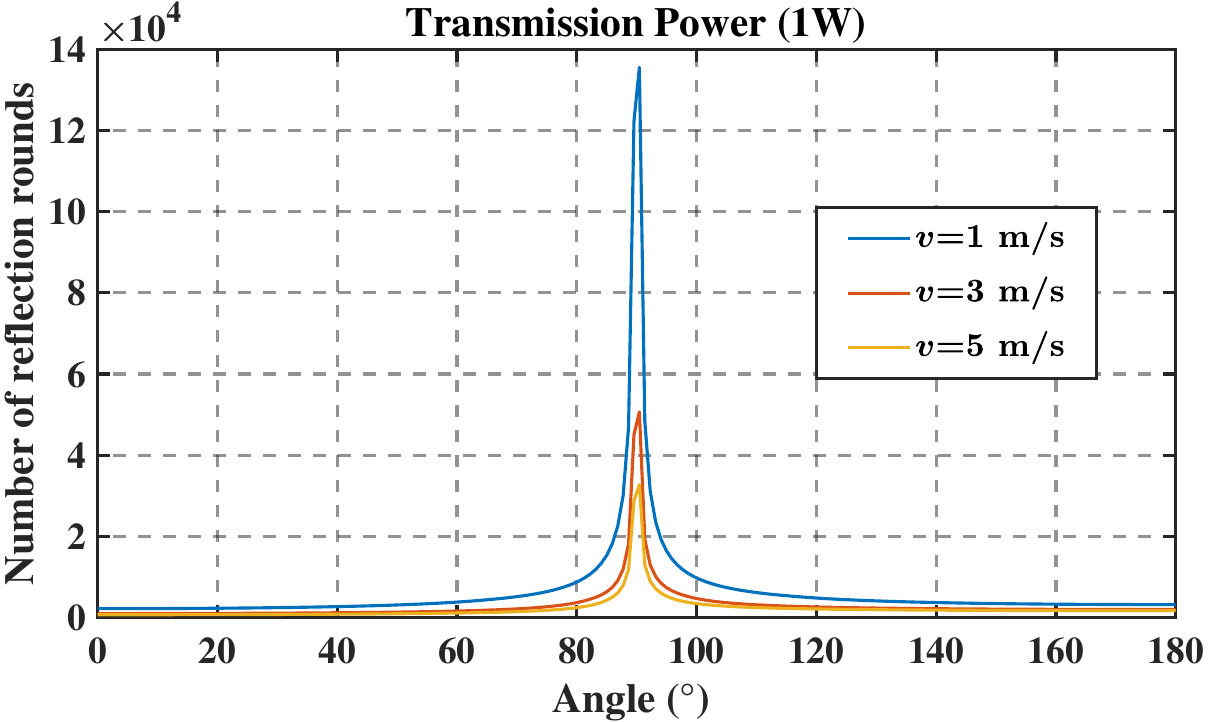}
   	\caption{Number of reflection rounds vs. direction angle under different speeds.}
   	\label{mobility}
   \end{figure}
  
   Fig. \ref{mobility} plots the number of reflection rounds as a function of direction angle under different speeds. From Fig. \ref{mobility}, we observe that the number of reflection rounds increases as the speed decreases, due to the slower increase of the accumulated Doppler shift at the lower speeds. Furthermore, the number of reflection rounds attains its maximum value at a direction angle of approximately 90$^{\circ}$, and rapidly decreases as the direction angle departs from 90$^{\circ}$, as a result of the accelerated growth of the accumulated Doppler shift as the direction angle departs from 90$^{\circ}$.

   In conclusion, we present a comparison of the mobility performance of the RBCom with that of VLC and FOC. For VLC, the utilization of LEDs with a large luminous angle enables the user to move freely within a specific range and stay connected to the access point (AP). Conversely, FOC exhibits relatively poor mobility due to the low response and high latency of the ATP system \cite{ATP}. The mobility performance of the RBCom is thus found to be better than that of FOC, but not as good as that of VLC.
    
   \subsection{Two-way Communication} 

   For VLC, signals are transmitted through LEDs and received through photo-diodes. The asymmetrical structure of the transmitter and receiver devices, along with the limitation imposed by the structure of LEDs on signal reception and demodulation, renders two-way communication unfeasible. For FOC, laser and photoelectric detectors are employed at the transmitter and the receiver for signal transmission and reception, respectively. The asymmetrical structure increases the difficulty of realizing two-way communication. In contrast, the RBCom exhibits good structural symmetry, which enables two-way communication through the addition of a set of demodulation devices at AP and a synchronization-based modulation at user equipment.

    \subsection{Safety} 

    For VLC, LEDs with low radiation energy refract at the surface of obstacles rather than penetrating through them. For FOC, the utilization of laser with high power and strong directivity poses a safety risk. In contrast, the RBCom exhibit improved safety feature. The presence of obstacles within the cavity breaks the resonance of the beam, causing the beam power to drop to zero within at most one reflection round. The energy produced during such a process is almost negligible and will not damage the obstacles, i.e, when the transmission power is 1 W and the communication distance is 10 m, the beam breaks in approximately 66.7 ns and produces 66.7 nJ of energy. Therefore, the RBCom and VLC have better safety than FOC.

    \subsection{Confidentiality}

     In WOC, the strong directivity plays a critical role in ensuring the confidentiality. LEDs with a large luminous angle result in the poor confidentiality of VLC, while laser with the strong directivity enables FOC to have high confidentiality. For RBCom, the resonant beam only propagates within the resonant cavity, and the self-alignment of the retroreflector ensures the strong directivity of the resonant beam. As such, the RBCom and FOC demonstrate enhanced confidentiality as compared to VLC.

    \section{{Future Directions and Challenges}}
    \label{Multi-user}
 
    This section first analyzes the underlying reasons contributing to the lower transmission rate of the RBCom compared that of FOC, and outlines potential avenues for improvement. Subsequently, a compensation method to prevent the termination under the mobile scenario is proposed. Finally, we introduce the basic framework of multiple access RBCom and highlight several implementation challenges that must be addressed in the future.

    \subsection{Improving Transmission Rate}

    As discussed in Section III-A, the modulation depth is limited by the gain capability of the gain mediums. Specifically, the saturation intensity imposes restrictions on the gain provided by the gain mediums\cite{Gainmeidum}, and if the modulation depth exceeds a certain threshold, the communication will be interrupted after a finite number of reflection rounds. Hence, it is imperative to maintain the modulation depth of the RBCom at a low level to ensure the stability of the communication link. Therefore, the future research directions of improving the transmission rate under the short and medium range scenarios mainly include the following two potential avenues. The first avenue entails exploring alternative information-bearing methods under the performance constraints of current devices, with the aim of simplifying the RBCom channel as much as possible while eliminating the echo interference. The second avenue focuses on the improvement of the performance of current devices. For instance, by reducing the response time of the pump source and the gain medium, the feasibility of the adaptive modulation scheme is improved, and a decrease in the saturation intensity of the gain medium leads to a corresponding increase in the modulation depth\cite{PartOne}.

    \subsection{Mobile Scenario}
      
    As discussed in Section III-B, the RBCom terminates after a certain number of reflection rounds. To conquer this issue, a frequency monitoring circuit is integrated into the synchronization module to detect the frequency of the beam in real-time. Additionally, a nonlinear optical crystal\cite{Compensation}, which converts the input beam frequency to the desired frequency through nonlinear optical interactions, is introduced into the cavity to generate a new resonant beam with the initial frequency when the frequency deviation is significant.

    \subsection{Multiple Access Scenario}

\begin{figure}[htb]
	\centering
	\includegraphics[scale=0.60]{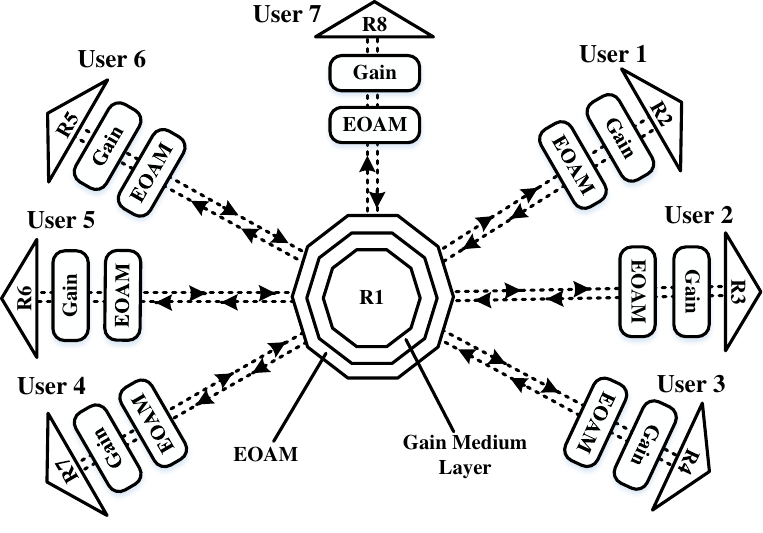}
	\caption{Basic framework of omnidirectional multiple access RBCom system.}
	\label{Architecture}
\end{figure}

   Fig. \ref{Architecture} shows the basic framework of the multiple access RBCom system composed of one AP and multiple users. At the AP, we use eight retroreflectors, each of which is with an effective range of 90$^{\circ}$, to splice the spherical retroreflector R1 covering the omnidirectional space, and place a gain medium layer outside R1 to form the resonant beam with the users. Additionally, an EOAM and a set of demodulation devices are placed in front of each retroreflector to modulate and demodulate the information symbols. Each user has a similar structure to the receiver shown in Section \ref{Structure}, using a retroreflector (i.e., R2$\sim$R8) to form a cavity with R1.
   
  Under this framework, users within the effective range of different retroreflectors are independent to each other. Therefore, we only focus on the scenario where all users are within the effective range of the same retroreflector. The uplink refers to a link from one or more users to the AP, where each user forms a cavity with the AP and the modulated beams sent by the users converge into one beam at the AP. The downlink refers to a link from the AP to one or more users, where the signals received by the users are identical due to modulation by the same EOAM.

  As discussed in Section \ref{Structure}, different distances between users and the AP result in different frame lengths for each user, which presents a challenge for realizing frame synchronization between the AP and users. To address this challenge in the implementation of the multiple access RBCom, a preliminary approach is proposed by redefining the frame length. Specifically, the frame length for each user is defined as the length of one reflection round with in its cavity, and the frame length for the access point is defined as the least common multiple of the frame lengths of all users. The AP and users then realize frame synchronization through the detection of the SS. However, further investigation into more efficient ways of realizing frame synchronization is deemed meritorious for future research.

  \begin{comment}
  Typical multiple access schemes including frequency-division multiple access (FDMA), space-division multiple access (SDMA), time-division multiple access (TDMA), and optical code-division multiple access (OCDMA) are introduced for the considered RBCom systems in the following.
  \end{comment}
  
  We take the frequency-division multiple access (FDMA) as a representative example to illustrate other implementation challenges of the multiple access RBCom systems. In this scheme, the whole frequency band of the gain medium is divided into several subbands to improve the resource utilization. Each user creates its own resonant beam with a different center frequency with the allocated subband. In the uplink, the AP receives a composite resonant beam consisting of beams with distinct center frequencies. The frequency division module is employed to decompose the received beam into individual resonant beams with distinct center frequencies. The AP then detects the signal of each user from the resonant beam with the corresponding center frequency. In the downlink, each user performs the same detection process as the AP. However, all detected signals are identical, since there exists only one EOAM at the AP. This poses a significant challenge for simultaneously modulating different signals onto different beams, which needs to be urgently addressed in future research.

\section{Concluding Remarks}
\label{Conclusion}

  In this article, a synchronization-based scheme was proposed to eliminate the echo interference in the RBCom, and the design for the transmitter and receiver was discussed. The proposed RBCom transceiver was compared with the conventional VLC and FOC in terms of transmission rate, mobility, two-way communication, and safety. Then, future research avenues for improving the transmission rate were outlined, and a compensation method to prevent the termination under the mobile scenario was proposed. Finally, the basic framework of the multiple access RBCom was introduced and several implementation challenges that must be addressed were highlighted.

\bibliographystyle{IEEEtran}   
\bibliography{resonant}

\begin{thebibliography}{10}
\providecommand{\url}[1]{#1}
\csname url@samestyle\endcsname
\providecommand{\newblock}{\relax}
\providecommand{\bibinfo}[2]{#2}
\providecommand{\BIBentrySTDinterwordspacing}{\spaceskip=0pt\relax}
\providecommand{\BIBentryALTinterwordstretchfactor}{4}
\providecommand{\BIBentryALTinterwordspacing}{\spaceskip=\fontdimen2\font plus
\BIBentryALTinterwordstretchfactor\fontdimen3\font minus
  \fontdimen4\font\relax}
\providecommand{\BIBforeignlanguage}[2]{{%
\expandafter\ifx\csname l@#1\endcsname\relax
\typeout{** WARNING: IEEEtran.bst: No hyphenation pattern has been}%
\typeout{** loaded for the language `#1'. Using the pattern for}%
\typeout{** the default language instead.}%
\else
\language=\csname l@#1\endcsname
\fi
#2}}
\providecommand{\BIBdecl}{\relax}
\BIBdecl

\bibitem{Survey-WOC}
M.~A. Khalighi and M.~Uysal, ``Survey on free space optical communication: A
  communication theory perspective,'' \emph{IEEE Commun. Surveys Tuts.},
  vol.~16, no.~4, pp. 2231--2258, 4th Quart., 2014.

\bibitem{LimVLC}
A.~Jovicic, J.~Li, and T.~Richardson, ``Visible light communication:
  Opportunities, challenges and the path to market,'' \emph{IEEE Commun. Mag.},
  vol.~51, no.~12, pp. 26--32, Dec. 2013.

\bibitem{WOC}
Z.~{Ghassemlooy}, S.~{Arnon}, M.~Uysal, Z.~Xu, and J.~Cheng, ``Emerging optical
  wireless communications-advances and challenges,'' \emph{IEEE J. Sel. Areas
  Commun.}, vol.~33, no.~9, pp. 1738--1749, Sep. 2015.

\bibitem{AppVLC}
J.~Shi \emph{et~al.}, ``3.76{-Gbps} yellow-light visible light communication
  system over 1.2 m free space transmission utilizing a {S}i-substrate {LED}
  and a cascaded pre-equalizer network,'' \emph{Opt. Express}, vol.~30, no.~18,
  pp. 33\,337--33\,352, Aug. 2022.

\bibitem{ATP}
Y.~{Kaymak} \emph{et~al.}, ``A survey on acquisition tracking and pointing
  mechanisms for mobile free-space optical communications,'' \emph{IEEE Commun.
  Surveys Tuts.}, vol.~20, no.~2, pp. 1104--1123, 2nd Quart., 2018.

\bibitem{AppFSO}
R.~Migliore, J.~Duncan, V.~Pulcino, D.~Bourne, S.~Voegt, and G.~Perez,
  ``Outlook on {EDRS-C},'' in \emph{Proc. Int. Conf. Space Opt. (ICSO)},
  Biarritz, France, Oct. 2016, pp. 105\,622S1--105\,622S9.

\bibitem{Resonant}
M.~Xiong, Q.~Liu, G.~Wang, G.~B. Giannakis, and C.~Huang, ``Resonant beam
  communications: Principles and designs,'' \emph{IEEE Commun. Mag.}, vol.~57,
  no.~10, pp. 34--39, Oct. 2019.

\bibitem{Echo}
M.~Xiong, Q.~Liu, G.~Wang, G.~B. Giannakis, S.~Zhang, J.~Zhu, and C.~Huang,
  ``Resonant beam communications with echo interference elimination,''
  \emph{IEEE Internet Things J.}, vol.~8, no.~4, pp. 2875--2885, Feb. 2021.

\bibitem{LongLaser}
G.~J. {Linford}, E.~R. {Peressini}, W.~R. {Sooy}, and M.~L. {Spaeth}, ``Very
  long lasers,'' \emph{Appl. Opt.}, vol.~13, no.~2, pp. 379--390, Feb. 1974.

\bibitem{Gainmeidum}
O.~Svelto, \emph{Principles of {L}asers}.\hskip 1em plus 0.5em minus
  0.4em\relax New York: Springer, 2005.

\bibitem{PartOne}
D.~Li, Y.~Tian, and C.~Huang, ``Design and performance of resonant beam
  communications{--}part {I}: the quasi-static scenario,'' \emph{IEEE Tran.
  Mobile Comput.}, Under Review.

\bibitem{EOAM}
T.~V. Schaijk, D.~Lenstra, K.~Williams, and E.~Bente, ``Model and experimental
  validation of a unidirectional phase modulator,'' \emph{Opt. Express},
  vol.~26, no.~25, pp. 32\,338--32\,403, Dec. 2018.

\bibitem{VLCdatarate}
D.~A. Basnayaka and H.~Haas, ``Hybrid {RF} and {VLC} systems: Improving user
  data rate performance of {VLC} systems,'' in \emph{Proc. IEEE Veh. Technol.
  Conf. (VTC)}, Glasgow, U.K., May 2015, pp. 1--5.

\bibitem{RBCcapacity}
D.~Li, Y.~Tian, and C.~Huang, ``Capacity analysis of mobile resonant beam
  communications,'' in \emph{Proc. IEEE Int. Conf. Commun}, Montreal, QC,
  Canada, June 2021, pp. 1--6.

\bibitem{Compensation}
P.~F. Bordui and M.~M. Fejer, ``Inorganic crystals for nonlinear optical
  frequency conversion,'' \emph{Annu. Rev. Mater. Sci.}, vol.~23, no.~1, pp.
  321--379, 1993.

\end{thebibliography}

\end{document}